# Piezoelectric Barium Titanate Nanostimulators for the Treatment of Glioblastoma Multiforme

*Attilio Marino[1,†,*], Enrico Almici[2,†], Simone Migliorin[2], Christos Tapeinos[1], Matteo Battaglini[1,3], Valentina Cappello[4], Marco Marchetti[5,6], Giuseppe de Vito[5,7], Riccardo Cicchi[5,7], Francesco Saverio Pavone[5,6,7], Gianni Ciofani[1,2,*]*

[1]Istituto Italiano di Tecnologia, Smart Bio-Interfaces, Viale Rinaldo Piaggio 34, 56025 Pontedera, Italy

[2]Politecnico di Torino, Department of Mechanical and Aerospace Engineering, Corso Duca degli Abruzzi 24, 10129 Torino, Italy

[3]Scuola Superiore Sant'Anna, The Biorobotics Institute, Viale Rinaldo Piaggio 34, 56025 Pontedera, Italy

[4]Istituto Italiano di Tecnologia, Center for Nanotechnology Innovation, Piazza San Silvestro 12, 56127 Pisa, Italy

[5]European Laboratory for Nonlinear Spectroscopy (LENS), Via Nello Carrara 1, 50019 Sesto Fiorentino, Italy

[6]Università di Firenze, Department of Physics and Astronomy, Via Giovanni Sansone 1, 50019 Sesto Fiorentino, Italy

[7]National Institute of Optics, National Research Council (INO-CNR), Largo Enrico Fermi 6, 50125 Firenze, Italy

[†]These authors contributed equally to this work

[*]CORRESPONDING AUTHORS:

attilio.marino@iit.it; gianni.ciofani@iit.it

ABSTRACT

Major obstacles to the successful treatment of gliolastoma multiforme are mostly related to the acquired resistance to chemotherapy drugs and, after surgery, to the cancer recurrence in correspondence of residual microscopic foci. As innovative anticancer approach, low-intensity electric stimulation represents a physical treatment able to reduce multidrug resistance of cancer and to induce remarkable anti-proliferative effects by interfering with $Ca^{2+}$ and $K^+$ homeostasis and by affecting the organization of the mitotic spindles. However, to preserve healthy cells, it is utterly important to direct the electric stimuli only to malignant cells. In this work, we propose a nanotechnological approach based on ultrasound-sensitive piezoelectric nanoparticles to remotely deliver electric stimulations to glioblastoma cells. Barium titanate nanoparticles (BTNPs) have been functionalized with an antibody against the transferrin receptor (TfR) in order to obtain the dual targeting of blood-brain barrier and of glioblastoma cells. The remote ultrasound-mediated piezo-stimulation allowed to significantly reduce *in vitro* the proliferation of glioblastoma cells and, when combined with a sub-toxic concentration of temozolomide, induced an increased sensitivity to the chemotherapy treatment and remarkable anti-proliferative and pro-apoptotic effects.

INTRODUCTION

Despite the dramatic efforts to develop diagnostic and therapeutic tools, the treatment of brain cancer remains a huge challenge in oncology, and successful treatments are still far from being attained. The main obstacles to the successful treatment of brain tumors include *i)* the structural complexity of the central nervous system, *ii)* the recurrence of the tumors, and *iii)* the acquired drug resistance during chemotherapy.[1] The most common and detrimental primary brain tumor among adults is represented by glioblastoma multiforme (GBM), a particularly aggressive malignant astrocytoma. Although various treatments are available for GBM, including surgical resection, chemotherapy, and radiation, prognosis remains extremely poor.[2] The average survival time following diagnosis of GBM patients is only fourteen months, while the five-year survival rate is about 5%.

As alternative anticancer approaches, effective physical treatments based on low-intensity alternating currents (AC) demonstrated great potential for inhibiting the proliferation of different kind of cancer cells without the use of any drug/chemical.[3,4] Specifically, AC is known to inhibit cell division by interfering with $Ca^{2+}$ and $K^+$ homeostasis and with the cytoskeletal components involved in cell division. Low-intensity AC resulted able to enhance the efficacy of a standard chemotherapy drug, temozolomide (TMZ), by reducing multidrug resistance,[5] and have been recently tested in combination with TMZ for the treatment of glioblastoma multiforme in clinical trials.[6,7] The involved mechanism seems to be mediated by a AC-dependent translocation of the drug transporter P-glycoprotein (P-gp) from the plasma membrane to the cytosol.[5] However, healthy brain cells (*i.e.*, human astrocytes) are also sensitive to AC-dependent antiproliferative effects[3] and, in this context, the local delivery of electrical stimuli to cancer cells is highly desirable.

The rapid development of innovative nanotechnological tools is allowing for the targeting of remote physical stimulations (*e.g.*, thermal, electrical, oxidative, ionic, *etc.*) in deep tissues.[8] In the field of nano-oncology, different nanotransducers have been designed to mediate photothermal, photodynamic, or magnetothermal conversion, and to locally deliver anticancer stimuli at tumor level.[9] These nanotechnology-assisted remote stimulation approaches exploit a non-invasive source of energy, such as, for example, alternated magnetic fields and near-infrared radiations, which penetrates the biological tissues and is finally transduced by the nanomaterial into another potential toxic form of energy (*e.g.*, heat).

In this context, our group proposed, for the first time in the literature, the remote electric stimulation of living cells mediated by piezoelectric nanoparticles,[10,11] an extremely interesting approach for the modulation of cell behavior and activities.[12,13] Taking advantage of the direct piezoelectric effect, these nanomaterials have been exploited to convert mechanical into electrical energy.[14,15] Electric potentials can be generated by piezoelectric nanoparticles in remote modality by using ultrasounds (US),[16] mechanical pressure waves that can be safely and efficiently conveyed into deep tissues. Electro-elastic mathematical models[11] allowed to estimate, at nanoparticle level, the magnitude of the output voltage ($\varphi_{output}$ ~ 0.5 mV) evoked in response to US intensity $I_{US} = 0.8$ W/cm$^2$, while electrophysiological recordings[17] and real-time Ca$^{2+}$/Na$^+$ imaging[11] of electrically excitable cells experimentally demonstrated the efficacy of nanoparticle-assisted piezo-stimulation. Recently, our group successfully exploited the antiproliferative effects of nanoparticle-assisted remote electric stimulation as non-invasive "wireless" therapy suitable for inhibiting proliferation of SK-BR3 breast cancer cells.[18] Similarly to low-intensity AC, chronic piezo-stimulations resulted able to inhibit cancer cell cycle progression by interfering with Ca$^{2+}$ homeostasis, by upregulating the gene expression of inward rectifier potassium channels, and by affecting the developing of the mitotic spindles during cell division.[18]

Associated to the difficulties of treatment of pathologies at the level of the central nervous system, we find the problem of blood-brain barrier (BBB) crossing. The recent growth of nanotechnology promises to revolutionize the delivery of nanomaterials across BBB to brain cancers.[19] At first instance, the delivery of different nanomaterials through the BBB at the tumor site can be efficiently obtained by taking advantage of the enhanced permeability and retention (EPR) effect.[20] This phenomenon is associated to a highly fenestrated and permeabilized BBB in correspondence of newly formed tumor vessels. A complementary strategy, that appears to be particularly relevant for diagnostic and therapeutic purposes, is the functionalization of nanomaterials with specific ligands to promote their BBB crossing and their targeting to specific cell types or anatomical districts.[21] Typical receptors on cancer cell membrane, as the folate, the transferrin, or the epidermal growth factor (EGF) receptors, can be targeted for an efficient delivery of nanostructures to cancer cells. Particular attention has been dedicated to the antibody against the transferrin receptor (anti-TfR Ab), since it can be successfully exploited as a dual-targeting ligand for both enabling the BBB-crossing and the uptake by cancer cells.[22-25]

In this work, we report the preparation of functionalized piezoelectric nanoparticles for *in vitro* BBB crossing, active glioblastoma cell targeting, imaging, and remote electric treatment. To this aim, tetragonal crystalline barium titanate nanoparticles (BTNPs) have been chosen as lead-free piezoelectric nanotransducers[26] because of their excellent level of biocompatibility,[27] high piezoelectric coefficient ($d_{33}$ ~ 30 pm/V),[28] peculiar optical properties,[29-31] and possibility to finely control their morphology.[32] Finally, the synergic effects of the chronic piezoelectric stimulation combined with sub-toxic TMZ treatment have been *in vitro* investigated.

MATERIALS AND METHODS

*Nanoparticle functionalization with anti-TfR antibody (AbBTNPs)*

Non-centrosymmetric piezoelectric barium titanate nanoparticles were purchased by Nanostructured & Amorphous Materials, Inc (nominal nanoparticle size 300 nm in diameter, as indicated by the provider, purity > 99.9%). In the literature, many different dispersing agents have been adopted to obtain a stable dispersion of these nanoparticles, like poly(vinylpyrrolidone) (PVP),[33] hexamethylenetetramine (HMT),[34] ascorbic acid,[35] and ethanolamine.[35] In this work, a wrapping with the amphiphilic 1,2-distearoyl-*sn*-glycero-3-phosphoethanolamine-N-[methoxy(polyethylene glycol)-5000] (DSPE-PEG, Nanocs, purity > 99%) was carried out, as this copolymer allows for easy and straightforward functionalization with many kinds of targeting moieties[18]. DSPE-PEG was mixed with BTNPs (1:1 w/w) in ddH$_2$O; the mixture underwent sonication with a tip sonicator (8 W for 150 s, Mini 20 Mandelin Sonoplus) and, after a centrifugation step (20 min at 900 rcf, Hettich®Universal 320/320R centrifuge), supernatant containing free DSPE-PEG was discharged. The wrapped nanoparticles were thereafter washed twice in ddH$_2$O and finally re-dispersed at a 5 mg/ml concentration in ddH$_2$O (for electron microscopy imaging), in PBS (for estimation of functionalization efficiency), or in complete cell medium (for stability studies and for biological experiments).

Concerning the nanoparticle functionalization with the antibody against the transferrin receptor, BTNPs were firstly coated with biotin-DSPE-PEG (20 % w/w, Nanocs, purity > 95%) and DSPE-PEG (80 % w/w), and subsequently conjugated to streptavidin-Ab anti-TfR (2.5 μg of Ab / mg of BTNPs, Abcore), similarly as described in a previous work.[18] Ab-functionalized BTNPs will be indicated in the text as AbBTNPs. DPSE-PEG-coated BTNPs have been used as control and will be indicated in the following as BTNPs for easiness of reading. The non-functionalized plain BTNP powder will be indicated as plain BTNPs. The quantification of Ab functionalization efficiency was carried out through the bicinchoninic acid (BCA) assay following the manufacturers' procedures (enhanced test tube protocol, Thermo Fisher). Nanoparticle size, Z-potential, and polydispersity of AbBTNP and BTNP suspension (100 μg/ml) were characterized by using dynamic light scattering (DLS, Nano Z-Sizer 90, Malvern Instrument); the dynamic measurements of size and polydispersity were performed every ten minutes for two hours. Fourier-transformed infrared spectroscopy (FT-IR) was performed using a Shimadzu Miracle 10 as previously described.[36]

*Multimodal imaging of BTNPs*

Imaging of BTNPs was performed by using scanning electron microscopy (SEM), second harmonic generation (SHG) microscopy, and confocal laser scanning microscopy (CLSM). A drop of the diluted BTNP dispersion (100 μg/ml) was deposited and let dry on a glass coverslip.

SHG imaging of tetragonal crystal lattice of piezoelectric BTNPs was carried out with a multimodal custom-made non-linear microscope using a femtosecond pulsed laser source (Discovery, Coherent Inc.) for excitation. Images were acquired using an excitation wavelength of 800 nm and a 20X water immersion objective lens (XLUM 20X 0.95 NA, Olympus Corporation). SHG signal at 400 nm was collected in the epi-direction using a dichroic filter. Emission spectrum was obtained exciting with a pump-and-probe beam at 810 nm and a Stokes beam at 1060 nm.

BTNPs were also detected by CLSM (C2s system, Nikon) with a 642 nm laser (emission collected at 670 nm $< \lambda_{em} <$ 750 nm), as showed elsewhere.[11,18,31] BTNP signal from SHG and CLSM images related to the same region of the glass coverslip were obtained and then merged with ImageJ software (https://imagej.nih.gov/ij/).

For SEM, the coverslip with the deposited nanoparticles was gold-sputtered at 60 nA for 25 s, and imaging was carried out by using a Helios NanoLab 600i FIB/SEM, FEI.

*Characterization of the blood-brain barrier model*

Cultures of immortalized brain-derived endothelioma bEnd.3 cell line (ATCC® CRL-2299™) were seeded at high confluence (seeding density $8\cdot10^4$ cells/cm$^2$) and maintained in proliferative conditions on 3 µm porous transwells (Corning Incorporated) in order to obtain a functional endothelial barrier mimicking the BBB.[37] In this configuration, endothelial layer separates the luminal compartment (on the top) from the abluminal compartment (on the bottom). Both the abluminal and luminal compartments were incubated with complete cell medium, composed by Dulbecco's Modified Eagle's Medium (DMEM, Thermo Fisher Scientific), 10% fetal bovine serum (FBS, Gibco), 100 IU/ml penicillin (Gibco), 100 µg/ml streptomycin (Gibco).

The development of a functional biological barrier was assessed at day 1, 3 and 6 of proliferation by measuring both FITC-dextran permeability and transendothelial electric resistance (TEER). BBB model permeability was analyzed by incubating the luminal compartment with 200 µg/ml of FITC-dextran (Sigma, molecular weight 70 KDa) and measuring the fluorescence emission ($\lambda_{ex}$ = 485 nm, $\lambda_{em}$ = 535 nm, Perkin Elmer Victor X3 UV-Vis spectrophotometer) of the abluminal compartment at different time points (10, 20, 30, 60, 120 min). TEER was assessed with a Millipore Millicell ERS-2 Volt-Ohmmeter device. Resistance across the plain transwell (blank) was subtracted to all the TEER measurements. After the quantitative BBB model characterizations, all the subsequent experiments reported in the text were performed on BBB models at day 3. The qualitative morphological integrity of the BBB models at day 3 and the expression of a specific marker of tight junctions (*zonula occludens-1*) were respectively verified by Coomassie® Brilliant Blue Staining (BioRad, 0.2% for 5 minutes) and by immunocytochemistry (please refer to the Materials and methods "Immunofluorescence staining").

*Investigations of nanoparticle-cell interactions and BBB model crossing*

BTNPs associated to bEnd.3 cells were observed with SEM imaging combined with energy-dispersive X-ray spectroscopy (EDX). Samples incubated for 30 min with 100 µg/ml BTNPs were washed twice in PBS and fixed with paraformaldehyde (PFA, 4 % in PBS). Subsequently, cells were washed twice with $ddH_2O$ and treated with glutaraldehyde solution (2.5 % in $ddH_2O$ for 30 min at 4°C) and dehydrated by using progressive ethanol gradients (0 %, 25 %, 50 %, 75 %, and 100 % in $ddH_2O$). Before SEM/EDX imaging (Helios NanoLab 600i FIB/SEM), samples were gold-sputtered as described above.

Concerning TEM imaging, samples incubated for 30 min with BTNPs or AbBTNPs were washed twice with PBS, fixed with a solution of 1.5 % glutaraldehyde in sodium cacodylate buffer (0.1 M, pH 7.4) and the pellet treated for epoxy resin embedding. Briefly, cells were post-fixed in 1% osmium tetroxide plus 1% $K_3Fe(CN)_6$ at room temperature; then cells were *en bloc* stained with 3 % solution of uranyl acetate in 20 % ethanol; finally, they were dehydrated and embedded in epoxy resin (Epon 812, Electron Microscopy Science). Polymerization has been performed for 48 h at 60°C. Samples were then sectioned with a UC7 Leica ultramicrotome equipped with a 45° diamond knife (DiATOME), and the slices of 80-90 nm were collected on 300 mesh copper grids. The ultrastructural analysis was performed by using a Zeiss Libra 120 Plus instrument operating at 120 kV equipped with an in-column omega filter.

Fluorescence staining of plasma membranes and acidic organelles in living bEnd.3 cells was carried out. For these experiments, bEnd.3 cells were seeded on 35 mm µ-dish (Ibidi) at $8·10^4$ cells/$cm^2$ density for 3 days and then incubated with 100 µg/ml BTNPs / AbBTNPs for 24 and 72

h. After nanoparticle treatment, cells were washed in PBS and stained with CellMask Green Plasma Membrane Stain (1:1000 dilution; Invitrogen) or with Lysotracker (50 nM; Invitrogen) following the manufacturers' procedures. Nuclear staining was performed with Hoechst 33342 (1 µg/ml, Invitrogen) in all samples. Finally, cells were washed and incubated with HEPES-supplemented (25 mM) phenol red-free DMEM (Thermo Fisher) supplemented with 10% of FBS for CLSM imaging (C2s system, Nikon). Images were acquired by using the same acquisition parameters for the different experimental classes and were subsequently analyzed with NIS-Elements software (Nikon). Concerning the analysis of nanoparticle internalization, signals of plasma membranes and nanoparticles were selected and then measured upon intensity thresholding. Intersections between the areas of BTNPs / AbBTNPs and plasma membranes or of nanoparticles and intracellular regions were then obtained and expressed as percentages of the total nanoparticle area. Co-localization between acidic organelles and nanoparticles was investigated by assessing Mander's overlap coefficient. 3D reconstruction of *z*-stack images was carried out by using NIS-Elements software (Nikon).

Investigations of nanoparticle internalization were also performed on U-87 cells (ATCC ® HTB-14), a cell line derived from a human primary glioblastoma that is well characterized and commonly used in brain cancer research.[38] The composition of the medium used for culturing U-87 cells was the same of that for bEnd.3 cells (U-87 seeding density $2 \cdot 10^4$ cells/cm$^2$). Internalization studies were performed by incubating 100 µg/ml of nanoparticles directly on U-87 cells seeded on 35 mm µ-dish (Ibidi). Alternatively, U-87 cells were seeded in the abluminal compartment of the transwell and 100 µg/ml of nanoparticles were dispersed in cell medium of the luminal compartment. Staining, CLSM imaging, and image analysis were carried out as described above for bEnd.3 cells. SHG imaging of nanoparticle internalization was carried out with the multimodal custom-made non-linear microscope described above, exploiting a pump-and-probe beam at 800 nm and a Stokes beam at 1040 nm. Coherent anti-Stokes Raman spectroscopy (CARS) signal at 650 nm and SHG signal from pump beam at 400 nm were acquired simultaneously in epi-direction.

BBB model-crossing was investigated through flow cytometry (CytoFLEX, Beckman Coulter). 100 µg/ml of nanoparticles were incubated in the luminal compartments of a BBB model. At 4 h, 24 h and 72 h of nanoparticle treatment, concentrations of BTNPs / AbBTNPs were assessed in the abluminal compartments. The number of events measured by flow cytometry was then converted to nanoparticle concentrations thanks to a calibration curve obtained at different known concentrations of BTNPs ($R^2$ = 0.997, Figure S1).

*Chronic ultrasound (US) stimulations and temozolomide (TMZ) treatment*

US were generated by a KTAC-4000 device (Sonidel) through a tip transducer (S-PW 3 mm diameter tip). Chronic US stimulations were applied with 1 W/cm$^2$ intensity and 1 MHz frequency. Single US stimuli lasted 200 ms and were delivered every 2 s, 1 h *per* day, for 4 days. This protocol of US treatment was chosen since was not able to detectably increase the temperature of the cell medium neither to affect cell behavior / proliferation.[17,18]

Concerning TMZ treatment, different concentrations of the drug (0-400 µg/ml) were assessed at 24 and 72 h in order to evaluate TMZ effects. The highest non-toxic concentration (50 µg/ml) was then tested in combination with US stimulations.

*Cell viability assays*

Metabolism of cell cultures after the treatment with temozolomide (TMZ, Sigma-Aldrich) and after chronic US stimulation was assessed with WST-1 Assay Reagent ((2-(4-iodophenyl)-3-(4-nitrophenyl)-5- (2,4-disulfophenyl)-2H-tetrazolium sodium salt, BioVision), as previously described.[39] Samples were washed twice with PBS and then incubated with the WST-1 reagent (1:10 dilution in complete medium with phenol red-free DMEM, 50 minutes at 37°C). The absorbance of the collected supernatants was measured with a multiplate reader (Perkin Elmer Victor X3 UV-Vis spectrophotometer); the blank, corresponding to the non-specific absorbance of the WST-1 dilution in phenol red-free DMEM, was subtracted from all measurements. Finally, all data were normalized with respect to the non-treated controls.

*Immunofluorescence staining*

Immunofluorescence was carried out to detect the expression of the tight junction marker *zonula occludens-1* (ZO-1) in the BBB model. PFA-fixed cells were incubated with a 0.1% Triton X-100 solution in PBS (25 min at room temperature) for membrane permeabilization and with 10% goat serum in PBS (1 h at room temperature) as a blocking solution. Subsequently, samples were treated with rabbit IgG primary antibody against ZO-1 (Invitrogen, 1:100 dilution in PBS supplemented with 10% goat serum, 3 h at room temperature) and, after 5 washes with PBS supplemented by 10% goat serum, were incubated with goat Alexa Fluor 488-IgG anti-rabbit secondary antibody (Invitrogen, 1:200 dilution in PBS supplemented with 10% goat serum, 2 h at room temperature). TRITC-conjugated phalloidin (100 μM, Millipore) and Hoechst 33342 (1 μg/ml, Invitrogen) were also included in solution with the secondary antibody in order to stain f-actin and nuclei, respectively.

Double immunofluorescence was performed to analyze the expression of the Ki-67 proliferation marker and of the p53 tumor suppressor marker on U-87 cells after 4 days of remote chronic piezoelectric stimulation and TMZ treatment. Immunocytochemistry was performed as described above with a primary mouse monoclonal anti-p53 antibody (Abcam, 1:200), a primary rabbit IgG anti-Ki-67 antibody (Millipore, 1:150), a TRITC-conjugated secondary anti-rabbit antibody (1:200, Millipore), and a FITC-conjugated secondary anti-mouse antibody (1:75, Millipore).

*$Ca^{2+}$ imaging*

$Ca^{2+}$ imaging was performed during US stimulation, with or without piezoelectric AbBTNPs, taking advantage of Fluo-4 AM $Ca^{2+}$-sensitive fluorescence dye, similarly as in a previous work.[11] Before US stimulation, U-87 cells were stained with Fluo-4 AM (Invitrogen, 1 µM in DMEM for 30 min at 37°C), washed twice with PBS and incubated with HEPES-supplemented (25 mM) phenol red-free DMEM (Thermo Fisher). Fluorescence time-lapse imaging was performed with CLSM (C2s system, Nikon), and obtained images were processed by using ImageJ (http://rsbweb.nih.gov/ij/). The average intracellular fluorescence intensity was defined as $F_0$ at time $t = 0$ s, and as $F$ for $t > 0$ s. $F/F_0$ values were calculated for both US and AbBTNPs+US experimental groups and reported in the graph.

*Statistics*

For multiple sample comparisons, ANOVA followed by Tukey's HSD *post-hoc* test was performed by using *R* software (https://www.r-project.org/); regarding the analysis of nanoparticle internalization in bEnd.3 and U-87 cells, independent two-sample *t*-tests were carried out by using Excel software. Statistically significant differences among distributions were indicated for $p < 0.05$. Finally, data were plotted in histograms as average ± standard error by using Excel software.

RESULTS

The scheme of the experimental design is represented in Figure 1. In Figure 1a the strategy of nanoparticle functionalization is depicted: piezoelectric BTNPs (showed in red) are wrapped with DSPE-PEG and DSPE-PEG-biotin, and subsequently conjugated with strepatavidin-Ab against human TfR to finally obtain AbBTNPs. In Figure 1b the luminal (in red) and abluminal (in light blue) compartments of the *in vitro* BBB model are showed, where bEnd.3 and U-87 cells are respectively seeded (cell membranes are shown in green, nuclei in blue, and AbBTNPs in red). After 72 h of nanoparticle incubation in the luminal compartment, U-87 cells exposed to nanoparticles that crossed the BBB model have been piezoelectrically stimulated with chronic US treatments, as schematically indicated in Figure 1c.

*Characterization and imaging of piezoelectric BTNPs*

Imaging of piezoelectric BTNPs is presented in Figure S2. Figure S2a reports a representative SEM image of the sample. Figure S2b represents the emission spectrum obtained by illuminating the tetragonal crystal lattice of piezoelectric BTNPs with a pair of spatially- and temporally-overlapped laser beams at 810 nm and 1060 nm (pump-and-probe beam and Stokes beam, respectively). Figure S2c shows the multi-modal imaging of piezoelectric BTNPs; signal of BTNPs observed by SHG of the pump beam (in red) co-localizes with that one detected by CLSM (in green).

Fourier transformed infrared spectroscopy (FT-IR) was performed in order to verify the successful functionalization of the BTNPs. Starting from the low wavelengths, the peaks in the range 530-600 $cm^{-1}$ (Figure S3) can be attributed to the Ti-O stretching bond and they are characteristic of the $BaTiO_3$ compound.[40] Shifting to higher wavelengths, the peak at 1450 $cm^{-1}$ that can be seen in spectrum *i*) of plain BTNPs can be attributed to an impurity of $BaCO_3$ as it has been reported elsewhere.[40] The peaks between 1000-1100 $cm^{-1}$ (spectrum *ii*)) are attributed to the C-O-C and C-O-H stretching[41] vibration of the aliphatic chain of poly(ethylene glycol) (PEG), while the peaks in the range 1600-1670 and 1300-1460 $cm^{-1}$ (spectrum *iii*)) can be attributed to the Amide I (C=O stretching)[42] and Amide III[40] vibrations of the attached anti-transferrin antibody. The peaks at 2280-2400 and 2850-3000 cm$^{-1}$ (spectra *ii*) and *iii*)) are attributed to the C-H stretching bond of the DSPE-PEG while the peak at 3320 cm$^{-1}$ (spectrum *iii*)) can be attributed both to the O-H stretching vibration of the DSPE-PEG/TfR antibody as well as to the Amide A (N-H stretching)[43] of the TfR antibody. The corresponding vibrations and wavelengths are summarized in Table S1. A small yet significant difference in Z-potential was observed between BTNPs (-29.6 ± 0.8 mV) and AbBTNPs (-22.0 ± 0.6 mV), thus further supporting the hypothesis of the successful functionalization of BTNPs with the Ab.

Quantitative measurements of functionalization efficiency indicated an amount of 1.1 ± 0.4 μg of Ab *per* mg of BTNPs (~ 44% of the Ab used for the reaction successfully linked to BTNPs). Considering the molecular weight of the Ab (~ 90 KDa) and the number of BTNPs *per* mg of powder ($1.2 \cdot 10^{10}$ particles / mg), about 624 ± 227 molecules of Ab were conjugated to each BTNP.

Polydyspersity index (PDI) and hydrodynamic diameter (*Rd*) of BTNPs and AbBTNPs were investigated (Figure S4). The PDI was found stable over time for both BTNPs and AbBTNPs (Figure S4a; 1 measurement / 10 min for 110 min total; for *t* = 0 min, 0.29 ± 0.05 for BTNPs and 0.25 ± 0.02 for AbBTNPs; for *t* = 110 min, 0.37 ± 0.04 for BTNPs and 0.37 ± 0.04 for AbBTNPs); in both cases 0.2 < PDI < 0.4, thus indicating a moderate dispersivity.[44] Furthermore, hydrodynamic diameter *Rd* of both samples were stable along the experiment (Figure S4b; 1 measurement / 10 min for 110 min total; for *t* = 0 min, *Rd* = 274 ± 1 nm for BTNPs and *Rd* = 252 ± 11 nm for AbBTNPs; for *t* = 110 min, *Rd* = 304 ± 13 nm for BTNPs and *Rd* = 280 ± 2 nm for AbBTNPs).

*AbBTNPs efficiently target endothelial-like cells and cross BBB model*

In order to obtain a functional biological barrier mimicking the BBB, confluent cultures of bEnd.3 cells were maintained in proliferative conditions on 3 μm porous transwell for 1, 3 and 6 days (BBB model characterization is reported in Figure S5). Transendothelial electric resistance (TEER) was measured to assess the performances of the barrier at the different time points (Figure S5a). After 3 and 6 days of maturation, BBB model showed similar TEER levels (41.9 ± 8.9 Ω·cm$^2$ and 48.5 ± 7.4 Ω·cm$^2$ at day 3 and day 6 of culture, respectively), in both cases significantly higher with respect to the 1 day culture (20.1 ± 0.9 Ω·cm$^2$; $p < 0.05$). The crossing of FITC-dextran through the BBB model (at day 1, 3 and 6 of maturation) is shown in Figure S5b and has been expressed as % of the maximum theoretically achievable fluorescence intensity in the abluminal compartment at different time points (10, 20, 30, 60 and 120 min). BBB model at day 3 and day 6 showed similar permeability to FITC-dextran after 20 min of incubation (19.8 ± 0.6% at day 3 and 17.8 ± 0.4% at day 6), while a significant lower permeability at day 6 of maturation (35.1 ± 1.7%) was observed after 120 min of dextran treatment with respect to both day 1 and day 3 (101.2 ± 2.7% at day 1 and 52.3 ± 2.4% at day 3; $p < 0.05$). The developing of cell multilayers was also observed at day 6. Considering the good performances of the BBB model at day 3 as well as the scarce mechanical stability and resistance to the shear forces of BBB immediately after day 6 (delamination and cell layer detachments were observed in different cultures starting from day 7-8), nanoparticle-crossing through barrier was tested on BBB starting from day 3. In Figure S5c the Coomassie (left image) and the immunofluorescence staining (right image, ZO-1 in green and nuclei in blue) of the 3-day BBB model are reported; it is possible to appreciate the complete maturation of functional junctions among bEnd.3 cells, that develop a endothelial layer separating the luminal from abluminal compartment of the transwell.

Analysis of BTNP / AbBTNPs interacting with bEnd.3 cells and assessment of BBB model crossing are shown in Figure 2. In Figure 2a the SEM imaging and the energy dispersive X-ray analysis (EDX) of BTNPs associated to the plasma membranes of bEnd.3 cells are reported. Qualitatively, TEM observations (Figure 2b) highlighted a higher amount of AbBTNPs associated to plasma membranes and up-taken by bEnd.3 cells with respect to the non-functionalized BTNPs. CLSM of immunofluorescence staining against the ZO-1 marker of tight junctions after 72 h of BTNP / AbBTNP treatment is showed in Figure 2c (nuclei in blue, f-actin in red, ZO-1 in green, nanoparticles in white). CLSM imaging revealed that both BTNPs and AbBTNPs were internalized in bEnd.3 cells; however, increased nanoparticle internalization can be appreciated in samples treated with AbBTNPs. Plasma membrane imaging was carried out at 72 h of nanoparticle treatment (Figure 2d) and showed a higher amount of nanoparticles internalized in cell body with respect to those associated to the plasma membranes (this was observed for both BTNPs and AbBTNPs). Histograms of Figure 2e and 2f show the cell membrane and intracellular areas (%) of bEnd.3 cells co-localizing with BTNPs / AbBTNPs at 24 and 72 h of nanoparticle incubation, respectively. The quantitative analysis demonstrates a significantly higher amount of AbBTNPs, both associated to membranes ($1.06 \pm 0.27\%$) and internalized by bEnd.3 cells ($2.04 \pm 0.30\%$), with respect to BTNPs ($0.30 \pm 0.14\%$ associated to plasma membranes and $0.55 \pm 0.31\%$ internalized in cells; $p < 0.05$) at 24 h. Furthermore, the amount of intracellular nanoparticles (both AbBTNPs and BTNPs) decreased from 24 to 72 h (AbBTNPs and BTNPs internalized in cells for 72 h correspond, respectively, to $1.02 \pm 0.14\%$ and $0.24 \pm 0.05\%$), likely due to the active transport mechanisms through the bEnd.3 cells (*e.g.*, transcytosis and exocytosis). Despite this decrement, the amount of functionalized nanoparticles internalized in bEnd.3 cells remained significantly higher with respect to BTNPs at 72 h ($p <$ 0.05). 3D reconstructions of nanoparticles (BTNPs / AbBTNPs in red) and bEnd.3 plasma membranes (in green) are available in Figure S6 (72 h of incubation). Moreover, co-localization analysis of nanoparticles and acidic cell compartments (*i.e.*, late endosomes and lysosomes) at 4, 24 and 72 h of BTNP / AbBTNP incubation is reported in Figure S7, and showed a progressive accumulation in the acidic organelles of the cells. Higher AbBTNPs co-localization with acidic cell compartment was found with respect to BTNPs at both 24 h (Mander's coefficients were $0.56 \pm 0.06$ for AbBTNPs and $0.36 \pm 0.03$ for BTNPs; $p < 0.05$) and 72 h (Mander's coefficients were $0.78 \pm 0.14$ for AbBTNPs and $0.36 \pm 0.07$ for BTNPs; $p < 0.05$), presumably as a consequence of the higher internalization level with respect to the non-functionalized ones.

In order to measure BBB model-crossing, BTNP / AbBTNP were incubated in the luminal compartment of the BBB model and nanoparticle concentrations in the abluminal compartment were measured at 4, 24 and 72 h of nanoparticle treatment. Results reported a progressive BBB-crossing of BTNPs / AbBTNPs at the different time points. Similar BTNP and AbBTNP concentrations were found at 4 h ($3.25 \pm 0.01$ µg/ml and $2.96 \pm 0.26$ µg/ml respectively for BTNP and AbBTNP) and 24 h ($6.26 \pm 0.83$ µg/ml and $6.72 \pm 0.10$ µg/ml respectively for BTNP and AbBTNP). Instead, a significantly higher BBB-crossing ability of AbBTNPs was observed at 72 h with respect to non-functionalized nanoparticles (~34% increase: $8.01 \pm 0.03$ µg/ml and $10.69 \pm 0.17$ µg/ml respectively for BTNP and AbBTNP; $p < 0.05$). All together, these results indicated a higher BBB-targeting and BBB-crossing efficiency of the functionalized nanosystem.

*Dual targeting of AbBTNPs*

Additionally to the measurements of nanoparticle concentration in the abluminal compartment, the ability of AbBTNPs to efficiently bind glioblastoma cells was tested (Figure 3). CLSM analysis of U-87 cells that were incubated for 24 h with 100 µg/ml BTNPs or AbBTNPs is shown in Figures 3a-b. Interestingly, a higher level of AbBTNPs (1.00 ± 0.23% intracellular nanoparticles and 1.32 ± 0.42% associated to membranes) were found with respect to BTNPs (0.16 ± 0.03% intracellular nanoparticles and 0.26 ± 0.11% associated to membranes; $p < 0.05$). Qualitative observations with CARS / SHG scans confirmed the higher level of AbBTNP internalization (Figure 3c). Moreover, the CLSM analysis of U-87 cells exposed to nanoparticles that crossed the BBB model was carried out; Figure 3d shows representative CLSM images of U-87 cells cultured in the abluminal compartment after a 72 h treatment with BTNPs or AbBTNPs in the luminal compartment. In Figure 3e, the quantitative co-localization analysis revealed a higher amount of AbBTNPs in the abluminal compartment that are associated to the plasma membranes (1.11 ± 0.35%) and internalized by U-87 cells (0.96 ± 0.25%) compared to BTNPs (0.38 ± 0.15% associated to plasma membranes and 0.16 ± 0.03%; internalized by U-87 cells; $p < 0.05$), thus demonstrating as the AbBTNPs resulted a successful nanosystem able to cross the *in vitro* BBB model and to target U-87 cells with higher efficiency with respect to the non-functionalized BTNPs. For this reason, the following experiments have been performed just by using AbBTNPs.

*Chronic piezoelectric stimulation inhibits proliferation of human glioblastoma cells*

Inhibitory effects of chronic piezoelectric stimulation on U-87 proliferation are shown in Figure 4. Two concentrations of AbBTNPs have been investigated: 100 µg/ml, already successfully tested on SK-BR3 breast cancer cells,[18] and 10 µg/ml, the concentration of nanoparticles able to cross the BBB model after 72 h. The expression of the nuclear proliferation marker Ki-67 has been analyzed through immunofluorescence assays combined with CLSM imaging (Figure 4a). Qualitatively, it is possible to appreciate a lower Ki-67 expression in piezoelectrically-stimulated cells (AbBTNPs+US) with respect to the control cultures (non-stimulated and non-incubated controls, cells incubated with AbBTNPs but non-stimulated with US, cultures stimulated with US without the presence of AbBTNPs); quantitative analysis of Ki-67$^+$ nuclei (%) are presented in Figure 4b and reported as the lowest proliferation rate was found for AbBTNPs+US cultures incubated with 100 µg/ml of nanoparticles (28.7 ± 2.5%), followed by AbBTNPs+US cultures incubated with 10 µg/ml of nanoparticles (51.1 ± 4.5%). Both these 2 experimental conditions resulted characterized by a significantly lower proliferation rate with respect to all the other control groups (72.3 ± 3.7% for non-stimulated and non-incubated controls, 77.4 ± 3.2% for cells incubated with AbBTNPs but non-stimulated with US, 86.8 ± 1.9% for cultures stimulated with US without the presence of AbBTNPs; $p < 0.05$).

Time-lapse $Ca^{2+}$ imaging on AbBTNPs+US (10 µg/ml) cultures demonstrated the successful remote activation of the cells (Figure 4c): remarkable long-term $Ca^{2+}$ waves are observed in response to the US stimulation in the presence of the nanoparticles. The peak of the $Ca^{2+}$ wave was detected ~ 5 min after starting the US stimulations, and the $Ca^{2+}$ concentrations remain higher than the basal levels even after 25 min of stimulation. The time lapses video of $Ca^{2+}$ imaging performed on US and AbBTNPs+US cultures are available as Supplementary Information (Video S1 and S2, respectively). The stability of $Ca^{2+}$ levels and the regular proliferation rate observed in response to the plain US stimulation (*i.e.*, without the presence of AbBTNPs) support the safeness of the proposed stimulation method.

*Synergic efficacy of remote piezoelectric stimulation with temozolomide treatment*

The ability of nanoparticle-assisted piezoelectric stimulation to improve the anticancer efficacy of temozolomide (TMZ) treatment was investigated (Figure 5). Toxic effects of TMZ were assessed by testing different concentrations of the chemotherapy treatment (0-400 µg/ml) at two different time points (24 and 72 h) through WST-1 assay (data are normalized and expressed as percentage of WST-1 absorbance values measured at 24 h on control cultures). Metabolism of U-87 cultures at 72 h of treatment was significantly affected when treating with concentrations at least of 200 μg/ml (Figure 5a): the best anti-proliferative effects were observed with the highest tested TMZ concentration (400 μg/ml), while first significant effects were observed by using 200 μg/ml. The hypothesis that piezoelectric stimulation could increase the sensitivity of TMZ was investigated by using 50 μg/ml of this chemotherapy drug, the highest concentration that was not effective in our testing conditions (Figure 5b-e). Experimental classes we represented by control cultures, cultures incubated with 10 μg/ml AbBTNPs, cultures incubated with 50 μg/ml TMZ, cultures incubated with 50 μg/ml TMZ and 10 μg/ml AbBTNPs, cultures chronically stimulated with US, cultures stimulated with US in the presence of 10 μg/ml AbBTNPs, cultures stimulated with US in the presence of 50 μg/ml TMZ, and, finally, cultures stimulated with US in the presence of 10 μg/ml AbBTNPs and of 50 μg/ml TMZ. WST-1 assay (Figure 5b) was performed at day 4 on control cultures (100.0 ± 7.2%), cultures incubated with AbBTNPs (101.3 ± 1.7%), cultures incubated with TMZ (97.4 ± 2.4%), cultures incubated with TMZ and 10 μg/ml AbBTNPs (95.3 ± 0.9%), cultures chronically stimulated with US (94.9 ± 4.5%), cultures stimulated with 10 μg/ml AbBTNPs+US (87.8 ± 1.3%), cultures stimulated with US and TMZ (94.7 ± 4.1%), and, finally, cultures stimulated with 10 μg/ml AbBTNPs+US in the presence of TMZ (TMZ+AbBTNPs+US; 72.1 ± 1.7%). Results confirmed the anti-proliferative effects of nanoparticle-assisted piezoelectric stimulation (AbBTNPs+US), that was able to significantly decrease the metabolic activity without the presence of TMZ with respect to the other control conditions (control, AbBTNPs, TMZ, AbBTNPs+TMZ, US, US+TMZ; $p < 0.05$). However, the major effects were observed by synergistically combining piezo-stimulation with TMZ (TMZ+AbBTNP+US; $p < 0.05$).

The expression of the Ki-67 proliferation marker and of the p53 tumor suppressor marker in response to 50 μg/ml TMZ, 10 μg/ml AbBTNPs+US, and of 10 μg/ml AbBTNPs+US with 50 μg/ml TMZ (TMZ+AbBTNPs+US) were compared with control cultures and are showed in Figure 5c. Qualitatively, a decreased number of cells and a lower Ki-67 expression were found in both AbBTNP+US and TMZ+AbBTNP+US experimental classes, compared to both control and TMZ. This observation is in line with the lowest metabolism levels reported in response to these treatments. Moreover, a higher amount of $p53^+$ nuclei was detected in response to TMZ+AbBTNP+US treatment with respect to the other experimental groups. Quantitatively, Ki-$67^+$ nuclei in control (72.4 ± 2.8%) and in TMZ-treated (65.8 ± 5.7%) cultures were significantly higher with respect to the cultures treated with AbBTNP+US (49.2 ± 3.7%; $p < 0.05$) and with TMZ+AbBTNP+US (27.7 ± 2.5%; $p < 0.05$), the last of which resulted the strongest antiproliferative treatment ($p < 0.05$; Figure 5d). Higher levels of $p53^+$ nuclei were found in response to the combined TMZ+AbBTNP+US therapy (28.3 ± 6.6%; $p < 0.05$; Figure 5e) with respect to all the other treatments (1.2 ± 1.3% for AbBTNP+US; 3.4 ± 1.1% for TMZ) and control cultures (1.0 ± 0.7%).

Overall, these results indicate that piezoelectric stimulation affects proliferation of U-87 cells and increases their sensitivity to TMZ. Indeed, TMZ therapy at non-toxic concentrations, when combined with chronic piezoelectric treatment, was able to promote cell apoptosis and reducing cell proliferation.

DISCUSSION

Recent advances in nanobiotechnology are directed to the development of smart and biocompatible sensors / actuators that are able to detect and respond to specific physicochemical conditions in the human body.[45-47] Piezoelectric nanomaterials are a promising class of nanostructures, that have been successfully exploited both as mechanical sensors for energy-harvesting and mechanobiology studies, and as nanostimulators for indirect electrical activation of excitable cells.[48,49]

In this work, we report for the first time the successful crossing of a piezoelectric nanomaterial through a BBB model. Piezoelectric barium titanate nanoparticles used in this study are characterized by a 300 nm diameter size, and resulted able to cross a BBB model with a quite good efficiency; crossing was however improved of ~30% by promoting nanoparticle targeting to BBB cells thanks to surface functionalization with anti-TfR Ab. These results are in line with observations of Wohlfart *et al.*, that reviewed various nanoparticles adopted for the delivery of different drugs into the brain and reported as most of the successfully ones are characterized by a size ranging from 150 to 300 nm.[50] Moreover, nanoparticles of 300 nm size are still small enough to passively cross the large defenestrations of the tumor-associated vessels developed during aberrant angiogenesis.[11,51] Indeed, the cutoff size of porous blood vessels in most of tumors is 380-780 nm, and 400 nm size nanoparticles are known to efficiently accumulate in the brain tumors.[52] These considerations are extremely important in view of exploiting piezoelectric BTNPs for *in vivo* and preclinical studies, especially considering the potential impact of these nanomaterials in nanomedicine, not only for brain cancer treatment, yet also for the non-invasive electric deep brain treatment of different neurodegenerative pathologies that are characterized by a defenestrated vasculature, such as Parkinson's and Alzheimer's diseases.[53]

The higher levels of AbBTNPs associated to plasma membranes and internalized in cell body with respect to BTNPs confirm the efficacy of the dual targeting strategy mediated by the antibody against TfR, a receptor highly expressed by the endothelial cells of the neurovascolature[54] and by different cancer cells (*i.e.*, glioma, lymphoma, leukemia, breast, lung, bladder).[55,56] In agreement with our observations, Cui *et al.* and Chang *et al.* exploited TfR targeting to promote the targeting of poly(lactic-co-glycolic acid) (PLGA) nanoparticles to glioblastoma cells, both *in vitro* and *in vivo*. In these cited works, PLGA nanoparticles were functionalized with Tf. However, recent researches reported a decrease of specificity of Tf-functionalized nanosystems in biological environment due to the high levels of endogenous free Tf.[57] Therefore, following an approach adopted also by other groups,[58,59] we performed nanoparticle functionalization with anti-TfR Ab, that does not compete with endogenous Tf for TfR binding. The cell-targeting efficiency of our nanoplatform was investigated by exploiting different imaging techniques, as SEM/EDX, TEM, CLSM and SHG, the last of which represents an advanced imaging technique allowing detecting the crystal asymmetry of BTNP tetragonal lattice. Taking advantage of these imaging approaches, nanomaterial was detected in biological samples without the need of any kind of surface modification (*e.g.*, with fluorophore functionalization or quantum-dot decoration) that can potentially interfere with nanomaterial-cell interaction and with its internalization fate. In this regard, thanks to their peculiar optical properties, non-centrosymmetric BTNPs display a potential impact for cancer theranostics.[31]

Concerning piezoelectric stimulation, AbBTNP+US treatment resulted able to affect the proliferation of different types of cancer cells, thus suggesting a high versatility of this anticancer approach. Particularly, we observed a remarkable decrease of proliferative U-87 cells after 4 days of chronic piezo-stimulation (from $86.8 \pm 1.9\%$ of Ki-67$^+$ nuclei, observed in control cultures, to $28.7 \pm 2.5\%$ of Ki-67$^+$ nuclei, when stimulating with 100 µg/ml AbBTNPs+US); the decrease in U-87 cell proliferation in response to piezoelectric stimulation was even more pronounced with respect to that observed on SK-BR-3 cells (from $80 \pm 8\%$ of Ki-67$^+$ nuclei observed in control cultures to $56 \pm 13\%$ of Ki-67$^+$ nuclei when stimulating with 100 µg/ml AbBTNPs+US).

Moreover, the anti-proliferative effects of piezoelectric stimulation resulted preserved, albeit to a lesser extent, when reducing nanoparticle concentration to 10 µg/ml (corresponding to the concentration of nanoparticles that crossed the BBB model after 72 h). No significant increases of apoptotic glioblastoma cells were observed when treating cells only with the piezoelectric stimulation. Instead, the piezo-stimulation approach, when combined with sub-toxic TMZ treatment, was able to significantly increase the percentage of apoptotic cells of about 25% and to further reduce the proliferation rate of the cells with respect to the piezo-stimulation alone. These results demonstrated as the nanoparticle-assisted remote piezoelectric stimulation increases the sensitivity of glioblastoma cells to TMZ treatment. The synergic attack (chemical, thanks to the chemotherapy drug, and physical, thanks to the remote electric stimulation) remarkably reduced the cell number and the metabolic activity of glioblastoma cultures. The remote piezo-stimulation has the potential to improve the therapeutic success by overcoming the main obstacles for brain tumor treatment indicated in the Introduction, and will be tested in more complex *in vivo* models and in preclinical studies. Particularly intriguing is the future perspective to target also small microscopic foci of the GBM, that are the main cause of the recurrence of the disease. A further point worth of investigation will be the analysis of the effects of nanoparticle size / morphology on anticancer effects; indeed, the size / shape also affect the values of piezoelectric and dielectric susceptibility coefficients,[60] thus not allowing to easily and independently control the nanomaterial morphology and its piezoelectric behavior.

Summarizing, the main novelties of this research consist in the preparation of piezoelectric nanoceramics able to cross a BBB model, to target glioblastoma cells, and to provide remote electric stimulations for increasing GBM sensitivity to TMZ-based chemotherapy; on the other hand, it is also necessary to underline the limits characterizing the present work, where *in vitro* models of BBB and GBM were adopted, thus highlighting once more the needing for future *in vivo* experiments.

CONCLUSIONS

We presented for the first time the preparation of functionalized piezoelectric BTNPs for BBB-crossing, active cancer cell targeting, imaging, and remote US-driven electric treatment. Moreover, we demonstrated the versatility of this nanotechnological approach, that allows the successful delivery of antiproliferative stimuli to glioblastoma cells. Furthermore, the chronic piezoelectric stimulation, in synergic combination with a sub-toxic concentration of TMZ, induced an increased sensitivity to chemotherapy treatment and remarkable anticancer effects.

All together, these findings open new interesting perspectives in nanomedicine, with a potential positive impact for the remote therapy of brain cancer and neurodegenerative conditions. Future works will be focused on investigating the efficacy of nanoparticle-assisted piezo-stimulation in xenograft models, in order to explore the realistic translation of these nanomaterials in the future clinical practice. The possibility to fabricate piezoelectric BTNPs with different size and higher piezoelectric coefficient by maintaining the same level of biocompatibility will be assessed, and the effects of nanoparticle morphology on BBB crossing and on piezo-stimulation efficiency will be evaluated. Moreover, the anticancer performances of remote piezo-stimulation approach will be tested in combination with TMZ for the treatment of TMZ-resistant glioblastoma cells, analyzing the molecular mechanisms at the base of TMZ resistance and sensitivity. Finally, the combination of piezo-stimulation with different anticancer drugs, radiotherapy and hyperthermia is envisaged in order to develop an efficient anticancer protocol for pre-clinical studies.

AUTHOR STATEMENT OF CONTRIBUTIONS

A.M. performed the nanomaterial functionalization, the piezoelectric stimulation experiments and the $Ca^{2+}$ imaging and contributed to write the manuscript. E.A. carried out the tests concerning the BBB crossing, and performed immunochemistry and statistical analysis. S.M. performed viability studies and CLSM imaging. C.T. carried out nanomaterial characterization (analysis of nanoparticle size, stability and BCA assay to evaluate the efficiency of nanomaterial functionalization). M.B. developed and characterized the BBB model (TEER analysis, ZO-1 expression and Coomassie® Brilliant Blue Staining). V.C. performed TEM analysis. F.S.P. and R.C. supervised the SHG acquisition. R.C. and M.M. built the SHG setup. M.M. and G.d.V. performed the SHG acquisition and analyzed SHG data. G.C. supervised and planned the whole work and contributed to write the manuscript. All authors have given approval to the final version of the manuscript.

ACKNOWLEDGEMENTS

This work has received funding from the European Research Council (ERC) under the European Union's Horizon 2020 research and innovation program (grant agreement N°709613, SLaMM).

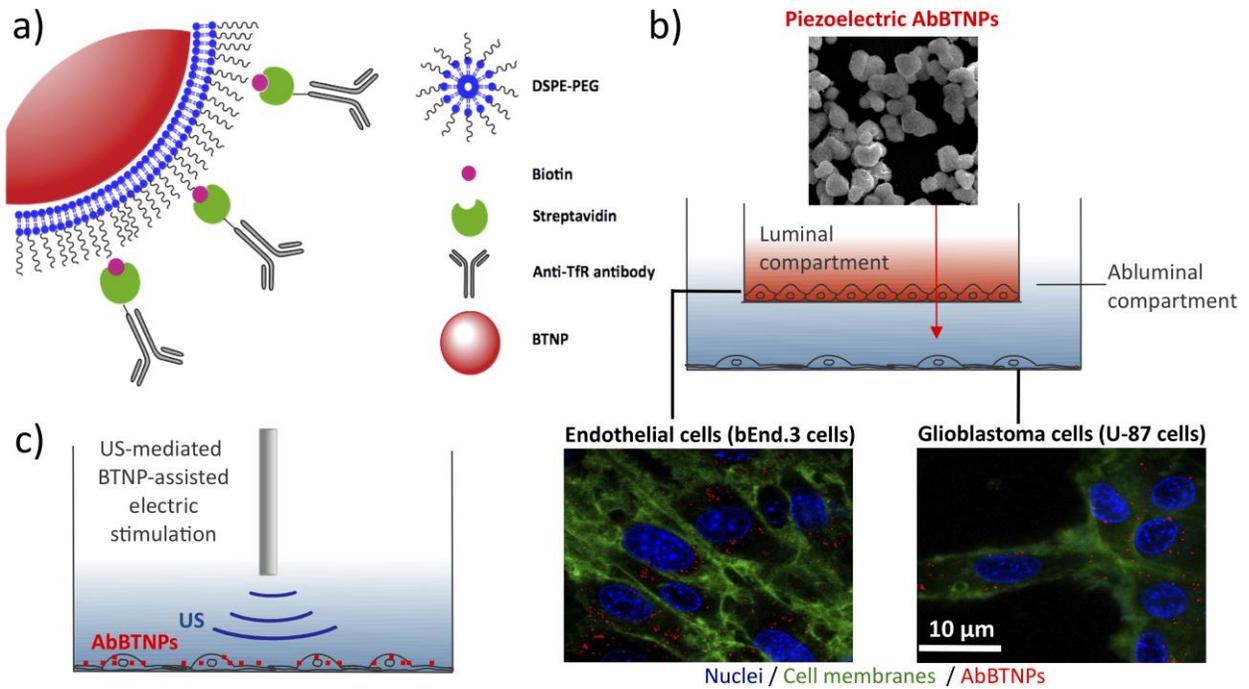

**Figure 1.** Experimental scheme of a) BTNP functionalization with antibody against transferrin receptor (TfR), b) nanoparticle crossing through a static 2D model of the BBB (nuclei in blue, cell membranes in green and AbBTNPs in red), and c) chronic piezoelectric stimulation of glioblastoma cells.

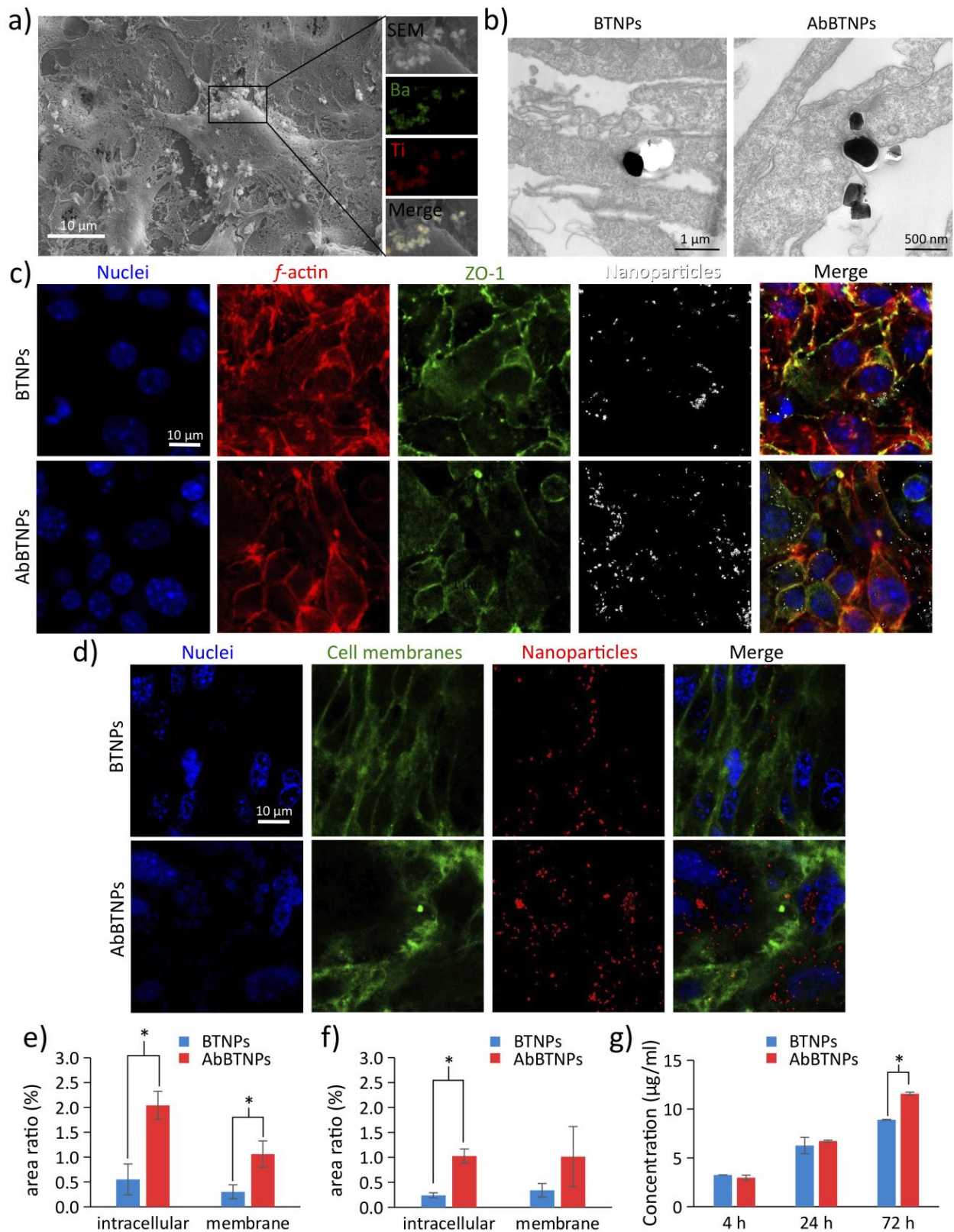

**Figure 2.** Analysis of BTNPs and AbBTNPs interaction with bEnd.3 cells and assessment of the BBB model crossing. a) SEM imaging and EDX analysis of BTNPs associated to the plasma

membranes of bEnd.3 cells (Ba in green and Ti in red). b) TEM image highlighting a higher amount of AbBTNPs associated to plasma membranes and up-taken by bEnd.3 cells with respect to the non-functionalized BTNPs. c) CLSM of immunofluorescence staining of bEnd.3 cells against the ZO-1 marker after 72 h of BTNP / ABTNP treatment (nuclei in blue, f-actin in red, ZO-1 in green and nanoparticles in white). d) CLSM imaging of bEnd.3 plasma membranes (in green), nanoparticles (in red) and nuclei (in blue), after 72 h of nanoparticle treatment. e-f) Histograms reporting intracellular and cell membrane areas (%) co-localizing with BTNPs / AbBTNPs after 24 and 72 h of nanoparticle incubation, respectively. g) Concentrations of BTNPs / AbBTNPs measured in the abluminal compartment after BBB crossing at different time points (4, 24 and 72 h). * $p < 0.05$.

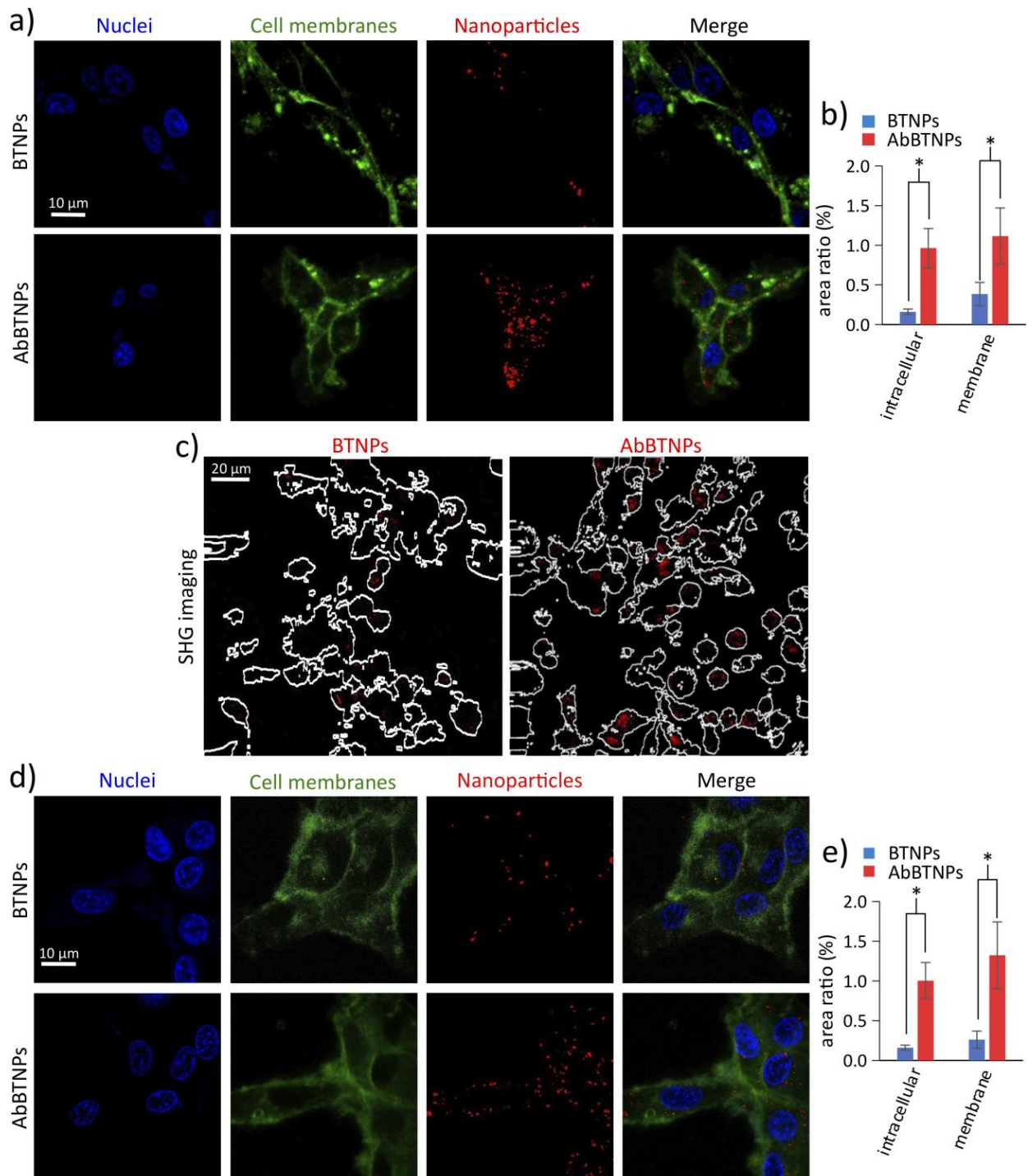

**Figure 3.** BTNP / AbBTNP targeting to glioblastoma cells. a-c) U-87 cells incubated for 24 h with 100 μg/ml BTNPs or AbBTNPs. a) CLSM imaging (plasma membranes in green, nanoparticles in red and nuclei in blue), b) histogram of nanoparticle localization, and c) SHG signal from nanoparticles (in red) overlaid on the outlines of the cells generated from the CARS images. d-e) CLSM analysis of U-87 cells exposed to nanoparticles after BBB model crossing, d) representative CLSM images of U-87 cells cultured in the abluminal compartment after 72 h of BTNP or AbBTNP treatment in the luminal compartment; e) quantitative analysis of experiment depicted in d). * $p < 0.05$.

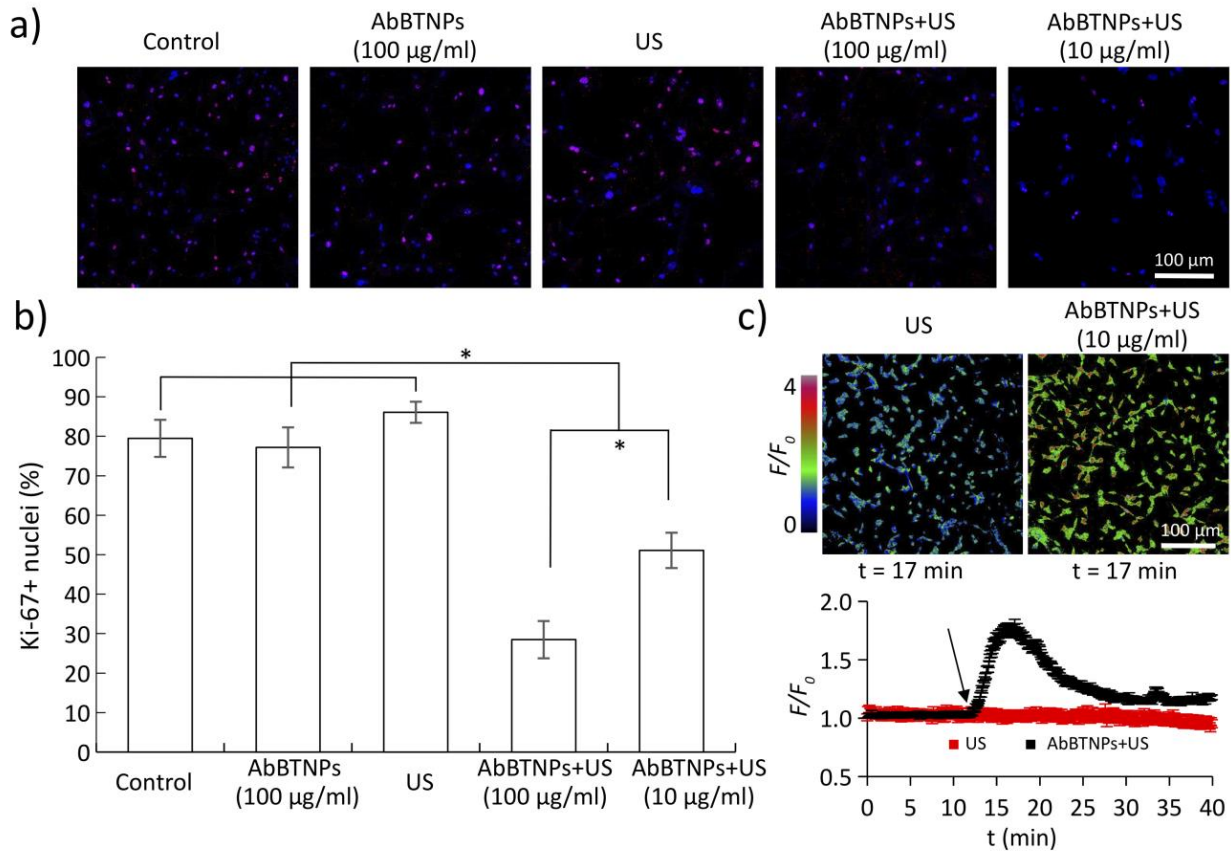

**Figure 4.** Inhibitory effects of chronic piezoelectric stimulation on U-87 proliferation by testing different AbBTNP concentrations (100 µg/ml and of 10 µg/ml, the latter corresponding to the concentration of nanoparticles crossing the BBB model after a 72 h treatment). a) CLSM analysis of Ki-67 proliferation marker on control cultures, AbBTNPs-treated cells, US-stimulated cells, and on AbBTNPs+US treated cultures. b) Histogram reporting Ki-67$^+$ nuclei (%). c) Time-lapse Ca$^{2+}$ imaging in response to plain US and to AbBTNPs+US (10 µg/ml). Images at the top show $F/F_0$ signal of cells after 5 min of US (top left) and AbBTNP+US (top right) stimulations. At the bottom, the graph reports $F/F_0$ traces of cultures stimulated with US (in red) and with US+AbBTNP (in black). * $p < 0.05$.

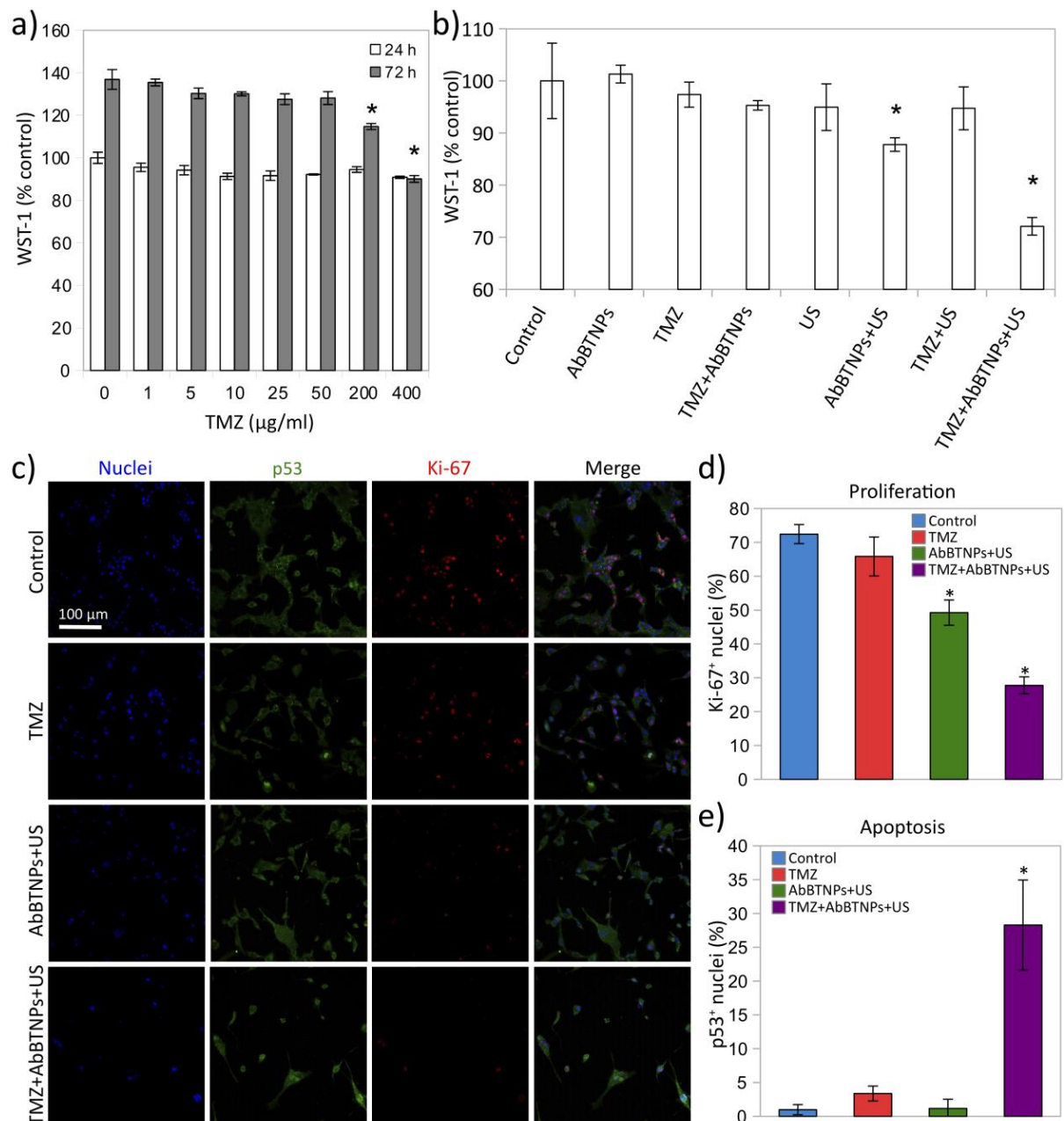

**Figure 5.** Nanoparticle-assisted piezoelectric stimulation (AbBTNPs+US) improves anticancer efficacy of temozolomide (TMZ). a) WST-1 assay on U-87 cells incubated for 24 h and 72 h with different concentrations of drug (0-400 μg/ml; data are normalized and expressed as percentage of WST-1 absorbance values measured at 24 h on control cultures). b) WST-1 assay respectively performed on control cultures, cultures incubated with 10 μg/ml AbBTNPs, cultures incubated with 50 μg/ml TMZ, cultures incubated with 50 μg/ml TMZ and 10 μg/ml AbBTNPs, cultures chronically stimulated with US, cultures stimulated with US in the presence of 10 μg/ml AbBTNPs, cultures stimulated with US in the presence of 50 μg/ml TMZ, and, finally, cultures stimulated with US in the presence of 10 μg/ml AbBTNPs and of 50 μg/ml TMZ. c) CLSM imaging of Ki-67 and p53 expression in the different experimental conditions. The histograms reporting the Ki-67$^+$ and p53$^+$ nuclei are respectively showed in d) and e). * $p < 0.05$.

## Graphical abstract

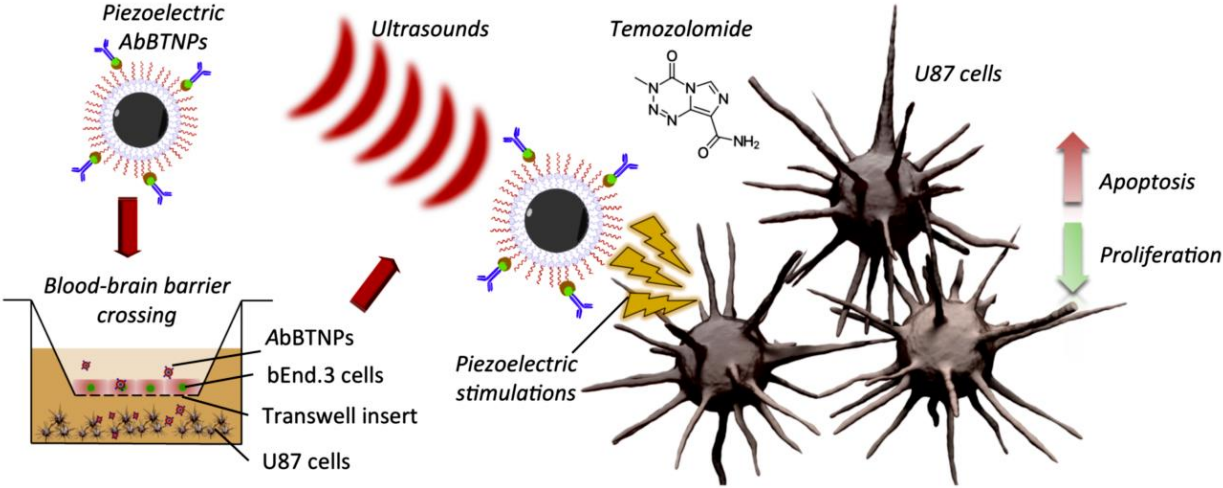